# Access control in a distributed micro-cloud environment

Tamara Ranković [1[0000-0002-5265-4626]] and Miloš Simić [2[0000-0001-8646-1569]] and Milan Stojkov [3[0000-0002-0602-0606]] and Goran Sladić [4[0000-0002-0691-7392]]

[1] Faculty of Technical Sciences, Novi Sad, Serbia, tamara.rankovic@uns.ac.rs
[2] Faculty of Technical Sciences, Novi Sad, Serbia, milos.simic@uns.ac.rs
[3] Faculty of Technical Sciences, Novi Sad, Serbia, stojkovm@uns.ac.rs
[4] Faculty of Technical Sciences, Novi Sad, Serbia, sladicg@uns.ac.rs

**Abstract.** Proliferation of systems that generate enormous amounts of data and operate in real time has led researchers to rethink the current organization of the cloud. Many proposed solutions consist of a number of small data centers in the vicinity of data sources. That creates a highly complex environment, where strict access control is essential. Recommended access control models frequently belong to the Attribute-Based Access Control (ABAC) family. Flexibility and dynamic nature of these models come at the cost of high policy management complexity. In this paper, we explore whether the administrative overhead can be lowered with resource hierarchies. We propose an ABAC model that incorporates user and object hierarchies. We develop a policy engine that supports the model and present a distributed cloud use case. Findings in this paper suggest that resource hierarchies simplify the administration of ABAC models, which is a necessary step towards their further inclusion in real-world systems.

## 1 Introduction

The widespread adoption of Internet of Things (IoT) systems in increasingly many industries has introduced novel and more challenging requirements for cloud computing. IoT devices intrinsically generate tremendous amount of raw sensor data and heavily rely on cloud for storage and processing. Some of them even require real-time responses, which is often infeasible due to long distances from data centers [1]. A recurring proposal that addresses these problems is to distribute cloud resources to smaller data centers in places of higher demand [2]. For example, edge computing places an edge layer between the cloud and devices [1].

Edge-IoT collaboration has significant potential to transform and accelerate the growth of many industries [3]. It is estimated there will be billions of edge clients only in the next few years [1]. For these predictions to come to life, it is necessary to rethink all aspects of the traditional cloud and adapt them to the edge environment.

In this paper, we focus on the access control to edge resources. Coarse-grained models like Role-Based Access Control (RBAC) [4] are generally not applicable in cloud-enabled IoT scenarios [1]. Authorization decisions should be based on many contextual conditions, including user, resource, spatiotemporal and relationship context [5]. Access control policies have to be flexible enough to support the following five necessary components: who, when, where, what and how [1]. On the other hand, policy administration has to be fairly straightforward in order to avoid misconfigurations, as a few administrators may manage thousands of devices.

The most prevalent solution, both in literature and in industry, is to opt for some form of ABAC [6] because of its flexible and dynamic nature [7]. ABAC enables declaring fine-grained policies, but that comes at the cost of high policy-management complexity. Extensions of ABAC that address this problem primarily do so by introducing hierarchies that facilitate permission inheritance implicitly through attribute inheritance. Majority of hierarchical ABAC models include user and object group hierarchies [8, 9] or attribute hierarchies [9], but there is little work on direct user and object hierarchies. More research needs to be conducted in this area in order to make ABAC models expressive enough to intuitively emulate widely-adopted constructs like role hierarchies [10]. Only by doing this, ABAC cam be fully utilized in domains where its flexibility is essential.

In this paper, we want to explore the implications of user and object hierarchies in ABAC models. Our goal is to determine whether and how hierarchies can simplify policy administration. Additionally, we want to investigate if this extension is useful in edge computing scenarios. To accomplish that, we formally define a Resource Hierarchy ABAC (RH-ABAC) model, a general-purpose model that allows creating user and object hierarchies via dependency relationships. We develop a policy engine that supports RH-ABAC to demonstrate model's feasibility and present a micro-cloud platform use case to display model's usability in the edge computing environment.

Any significant increase in real-world systems that rely on ABAC will be possible only when the administrative overhead is reduced enough so that benefits outweigh the complexity. Our research presents one step toward that goal. Another contribution of our paper lies in the use case we present, as it can help researchers identify requirements for access control in edge computing systems.

The rest of the paper is organized as follows: Section 2 presents relevant literature review. In section 3, we present the RH-ABAC model: components, formal definitions and policy conflict resolution strategy. Section 4 outlines the policy engine architecture and implementation details. In section 5 we present a use case that demonstrates RH-ABAC usability in distributed micro clouds. Section 6 concludes the paper and discusses directions of future work.

## 2 Related work

User groups simplify policy administration as users inherit permissions from their groups [11]. RBAC extended this mechanism with roles, which group users and permissions simultaneously. Permissions are added and removed from roles, which affects all users assigned those roles. In $RBAC_1$ model [12], role hierarchies additionally reduce redundant permission assignments, as they enable roles to inherit permissions from their junior roles.

While user-role and role-role relationships clearly express inheritance in RBAC, permissions are vaguely described as operations on objects, without any guidance on how to structure and reason about them. This has led to additional development of RBAC models. For example, authors of [7] formalize the access control model employed by Google Cloud Platform (GCP), where permissions can be operations on objects or object types. If someone is allowed to perform an operation on an object type, they are also allowed to do it on all objects of that type.

Straightforward RBAC management comes at the cost of inflexible policies that are insufficient in many cases, when hybrid models taking attributes into account or pure ABAC models are seen as a preferable option [7, 10]. ABACα [6] is a formal model containing only a minimal set of ABAC elements: users, subjects, objects, user attributes, subject attributes, object attributes, policies based on attributes and constraints. Even though ABACα policies can be highly expressive and direct relationship representation via attributes is trivial, it is not easy to represent hierarchy structures, proven to be very useful for inheritance [10].

Hierarchical Group and Attribute-Based Access Control (HGABAC) model [8] attempts to solve this problem via user and object groups. Users can be assigned to groups from which they obtain attributes. If an attribute is assigned to a group, all members of that group inherit it. Additionally, groups form a group graph, which enables attribute propagation from parent groups to children groups. Object groups and attribute inheritance work identically. Authors concluded that HGABAC requires less operations to express certain access control rules when compared to ABACα, while for others it has an administrative overhead. Another potential problem is the authorization function, requiring only one policy to evaluate to true for a permission to be granted, which isn't necessarily desirable behavior. If one policy states that a group should have access to a resource and the other one denies access to a user from that group, the user would still have access because the of the first policy. Merging the two policies solves the problem, but it leads to unmaintainable policies as exceptions stack up.

A slight variation of HGABAC, restricted HGABAC (rHGABAC), is proposed in [9]. Authors altered the policy format in order to implement a policy engine backed by rHGABAC. Groups and group hierarchies remained unchanged. To further simplify attribute management process, rHGABAC model introduced attribute hierarchies. Attribute values form a partially ordered set. If a user is assigned a senior attributed value, it is also assigned all junior attribute values. For example, role and resource hierarchies can be represented in this way. Role hierarchy can be formed in the attribute value set and by assigning one value from the set, which is a role, all junior roles to that role would also be assigned. In this way, all users with senior roles would satisfy policies requiring junior roles.

Hierarchical ABAC models have also been developed in the area of IoT and edge computing. Access control model for a smart cars use case utilizing attribute inheritance is presented in [13]. Multiple objects, usually sensors and actuators, form clustered objects, for example smart cars. Clustered objects can be direct members of one group and groups can be organized hierarchically. Aside from attributes an object was directly assigned, it inherits attributes of its clustered object. Clustered objects inherit additional attributes from their direct group and a group inherits attributes from its senior groups. Another model, developed for cloud-enabled industrial IoT, was introduced in [3]. The group concept is divided into static and dynamic groups. In the case of static groups, members are explicitly declared, while members of dynamic groups are objects that satisfy the query set by an administrator. Aside from attribute inheritance, groups enable policy inheritance. Policy attached to a group is propagated to all children groups and object members.

## 3     RH-ABAC model

This section provides an overview of the RH-ABAC model and its core components, including sets, relations and authorization function. Fig. 1 shows core relations between the sets, while formal model definitions can be found in Table 1.

**Resources (R)** is a set consisting of all entities present in the system. It is a union of two sets with no overlapping values: **Users (U)** and **Objects (O)**. Users are entities that can perform operations, while objects are entities on which those operations are performed. Operations are not directly performed by users, but rather by their representations in the **Subjects (S)** set. Each subject is created by only one user and it executes actions on their behalf, but there can be multiple subjects that correspond to the same user. *User* function maps every subject to its creator. It is worth noting that subjects are equivalent to sessions in the RBAC model.

All resources, both users and objects, are assigned a set of attributes. They represent the basis on which fine-grained access control policies can be defined. Every attribute is a *(name, value)* pair that belongs to the **Resource Attributes ($A_R$)** set. Attribute assignments are defined by the **Resource Attribute Assignment ($A_RA$)** relation. A resource is constrained to have up to one attribute with the same name. Subjects obtain attributes from users by the *SubAttr* mapping. A subject can hold all attributes of its creator or only a subset of them. This enables the enforcement of the least privilege principle.

In addition to attributes inherited from the user, subjects can be assigned one or more attributes from the **Request Attributes ($A_{REQ}$)** set. Assignments are represented by the **Request Attribute Assignment ($A_{REQ}A$)** relation. Their purpose is to describe conditions at the time of the subject request arrival more closely. Conditions can be related to the environment (e.g. current time), user connection (e.g. the IP address the request originated from) or something else. Since attribute values can be different for each request a subject makes, they should be updated accordingly.

Operations that can be performed in the system are elements of the **Operations (OP)** set. An operation paired with a value from the **Permission Types (PT)** set forms an element in the **Permissions (PERM)** set. This allows for positive and negative permissions. In other words, it is possible to explicitly allow or deny access to a certain resource or group of resources. Permissions are combined with elements of the **Conditions (C)** set to create **Policies (POL)**. Each condition is a function that returns a Boolean value based on object's attributes and subject's resource and request attributes.

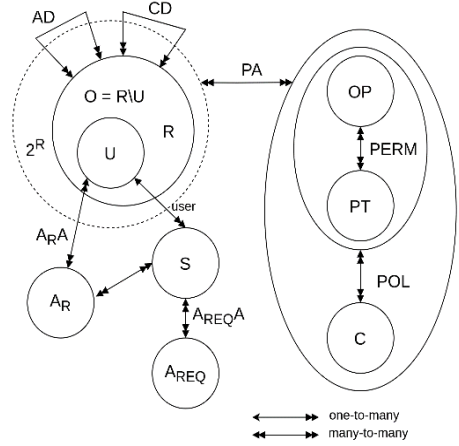

**Fig. 1.** RH-ABAC conceptual model

Subjects will be granted a permission *p* only if there is a policy that contains *p* and its condition evaluates to *true* for provided attribute values. Conditions should be specified in some policy language, which is outside the scope of this paper.

**Aggregation Dependencies (AD)** and **Composition Dependencies (CD)** are strict partial order relations on resources. They both express *part-of* relationships between resources. The difference between them is that the child's life depends on the parent's life cycle in the case of composition, but not in the case of aggregation. **Dependencies (D)** relation, also a strict partial order relation, is the union of aggregation and composition dependencies extended by pairs required to satisfy transitivity. Resources and dependencies form a **Resource Dependency Graph (RDG)**, a directed acyclic graph. These dependencies can be viewed as additional resource attributes, represented in an alternative way. For example, to emulate group memberships, there can be an aggregation dependency from the group to the user. It is equivalent to assigning the user a *group* attribute.

Since dependencies can't be referenced in policy conditions, as they are not regular attributes, there has to be a mechanism for accessing them. The problem is solved by introducing user and object scopes. They are additional conditions assigned to policies through **Policy Assignments (PA)**. Every element of the PA set is a *(policy, subject scope, object scope)* triple. Both the subject and object scope are non-empty sets of resources. For a permission to be granted, aside from the condition that must return true, the subject must be in the subject scope and the object must be in the object scope. A subject is in the subject scope if all resources from the scope are subject's parents, defined by the *SubParents* function, or they are equal to the user who created the subject. The same principle is applied to objects, with the difference that object's parents are determined by the *ResParents* mapping. Multiple resources in the scope correspond to the *AND* logical operator. For example, subject scope *{group: g1, group: g2}* states that a subject must be a member of both the group *g1* and group *g2*.

**Table 1.** RH-ABAC sets, relations and functions

| |
|---|
| **Sets and relations** |
| • $R, U(U \subseteq R), O(O = R \setminus U), S, A_R, A_{REQ}$ - sets of resources, users, objects, subjects, resource attributes and request attributes |
| • $OP, PT = \{allow, deny\}, PERM = OP \times PT$ - sets of operations, permission types, and permissions |
| • $C$ - the set of conditions, where $\forall c \in C$, c is $f: 2^{A_R} \times 2^{A_R} \times 2^{A_{REQ}} \longrightarrow \{true, false\}$ |
| • $POL \subseteq PERM \times C, AD = \{allowed, denied, undefined\}$ - sets of policies and authorization decisions |
| • $A_R A \subseteq R \times A_R$ - resource attribute assignment (many-to-many relation) |
| • $A_{REQ} A \subseteq S \times A_{REQ}$ - request attribute assignment (many-to-many relation) |
| • $PA \subseteq POL \times 2^R \times 2^R$ - policy assignment (ternary relation), where the combination of the operation, permission type, user scope and object scope must be unique, $\forall pa \in PA: pa = (op \in OP, pt \in PT, c \in C, user\_scope \in 2^R, object\_scope \in 2^R)$ |
| • $\preceq_{AD} \subseteq R \times R$ - aggregation dependency, a strict partial order relation on $R$ |
| • $\preceq_{CD} \subseteq R \times R$ - composition dependency, a strict partial order relation on $R$ |
| • $\preceq_{AD} \cap \preceq_{CD} = \emptyset$ • $r_1 \preceq_{AD} r_2 \Rightarrow r_2 \not\preceq_{CD} r_1$ |
| • $root \in R: (\forall r \in R)(r \preceq_{CD} root)$ |
| • $\preceq_D = \preceq_{AD} \cup \preceq_{CD} \cup \{(r_1, r_3) | (r_1, r_2) \in (\preceq_{AD} \cup \preceq_{CD}) \wedge (r_2, r_3) \in (\preceq_{AD} \cup \preceq_{CD})\}$ |
| • $RDG = (R, \preceq_D)$ - Resource Dependency Graph, a directed acyclic graph |
| **Functions** |
| • $User: S \longrightarrow U$ - mapping a subject to the corresponding user |
| • $ResAttr: R \longrightarrow 2^{A_R}, ResAttr(r) = \{a \in A_R | (r, a) \in A_R A\}$ - mapping a resource to its attributes |
| • $SubAttr: S \longrightarrow 2^{A_R}, SubAttr(s) \subseteq ResAttr(User(s))$ - mapping a subject to its attributes |
| • $ReqAttr: S \longrightarrow 2^{A_{REQ}}, ReqAttr(s) = \{a \in A_{REQ} | (s, a) \in A_{REQ} A\}$ - mapping a subject to request attributes |
| • $ResParents: R \longrightarrow 2^R, ResParents(r) = \{r_1 \in R | r \preceq_D r_1\}$ - mapping a resource to its parent resources |
| • $SubParents: S \longrightarrow 2^R, SubParents(s) \subseteq ResParents(User(s))\}$ - mapping a subject to its parent resources |
| • $ResChildren_C: R \longrightarrow 2^R, ResChildren_C(r) = \{r_1 \in R | r_1 \preceq_{CD} r\}$ - mapping a resource to its children in composition relation |
| • $TransitiveReduction: 2^R \times 2^{R \times R} \longrightarrow 2^R \times 2^{R \times R}$ - mapping a graph to its transitive reduction |
| **Policy evaluation** |
| $PolEval: POL \times 2^{A_R} \times 2^{A_R} \times 2^{A_{REQ}} \longrightarrow AD, PolEval(pol, SubAttrs, objAttrs, ReqAttrs) =$ $\begin{cases} allowed & pol_{(2)}(SubAttrs, objAttrs, ReqAttrs) = true \wedge pol_{(1)} = allow \\ denied & pol_{(2)}(SubAttrs, objAttrs, ReqAttrs) = true \wedge pol_{(1)} = deny \\ undefined & \text{otherwise} \end{cases}$ |
| $EffectivePolicies: S \times R \times OP \longrightarrow 2^{PA}, EffectivePolicies(sub, obj, op) = \{pa = (pol, US, OS) \in PA | (\forall us \in US)(us \in (SubParents(sub) \cup User(sub))) \wedge (\forall os \in OS)(os \in (ResParents(obj) \cup obj)) \wedge pol_{(0)} = op\}$ |
| $PolPriorityBySub: S \times PA \longrightarrow \mathbb{Z}, PolPriorityBySub(sub, pa) = max(\{p = -Dist(User(s), userScope) | userScope \in pa_{(3)}\})$ |

$PolPriorityByObj: O \times PA \rightarrow Z, PolPriorityByObj(obj, pa) = max(\{p = -Dist(o, objScope) | objScope \in pa_{(4)}\})$

• $Dist$ calculates distance between two nodes in the $TransitiveReduction(RDG)$ graph

$$MaxPriorityBySub: S \times O \times 2^{PA} \rightarrow 2^{PA},$$

$MaxPriorityBySub(sub, obj, pas) = \{pa \in pas | Eval(pa_{(0,1,2)}, SubAttr(sub), ResAttr(obj), ReqAttr(s)) \neq undefined \land \neg(\exists pa_1 \in pas)(pa_1 \neq pa \land Eval(pa_{1(0,1,2)}, SubAttr(sub), ResAttr(obj), ReqAttr(s)) \neq undefined \land PolPriorityBySub(sub, pa_1) > PolPriorityBySub(sub, pa))\}$

$$MaxPriorityByObj: S \times O \times 2^{PA} \rightarrow 2^{PA},$$

$MaxPriorityByObj(sub, obj, pas) = \{pa \in pas | Eval(pa_{(0,1,2)}, SubAttr(sub), ResAttr(obj), ReqAttr(s)) \neq undefined \land \neg(\exists pa_1 \in pas)(pa_1 \neq pa \land Eval(pa_{1(0,1,2)}, SubAttr(sub), ResAttr(obj), ReqAttr(s)) \neq undefined \land PolPriorityByObj(obj, pa_1) > PolPriorityByObj(obj, pa))\}$

$$Eval: S \times O \times 2^{PA} \rightarrow AD,$$

$Eval(sub, obj, pas) = \begin{cases} undefined & pas = \emptyset \\ denied & (\exists(pol, us, os) \in pas)\left(PolEval\begin{pmatrix} pol, SubAttr(sub), \\ ResAttr(obj), \\ ReqAttr(sub) \end{pmatrix} = denied\right) \\ allowed & otherwise \end{cases}$

$$Authorize: S \times O \times OP \rightarrow AD,$$

$Authorize(s, o, op) = Eval\left(MaxPriorityByObj\begin{pmatrix} s, o, \\ MaxPriorityBySub\begin{pmatrix} s, o, \\ EffectivePolicies(s, o, op) \end{pmatrix} \end{pmatrix}\right)$

Only policies present in the PA set are evaluated when an access control request is being processed. That poses the question of how to represent policies that should be applicable to all subjects or objects. The R set includes a *root* resource that is connected to all other resources by the composition relationship. When a policy should be evaluated for every subject, the policy's subject scope should contain only the root resource. The same strategy goes for objects.

Every authorization request contains a subject, object and operation and returns a value from the **Authorization Decisions (AD)** set. *EffectivePols* calculates policies that should be evaluated for a request, based on the operation and subject and object scopes. Selected policies can make conflicting decisions when evaluated by the *PolEval* function. For that reason, we present a conflict resolution strategy. Each policy has a subject priority, defined by *PolPriorityBySub* and object priority, defined by *PolPriorityByObj*. The greater the distance of the subject's creator from the subject scope in RDG, the smaller the policy's subject priority. Object priority is determined in the same way. Out of all applicable policies, only the ones that have a defined authorization decision and the maximum subject priority should be kept. From them, only those with the maximum object priority are selected. If there is any policy in the resulting set that evaluates to the *Denied* decision, the operation is denied. If the set is empty, the decision is undefined. In other cases, the operation is allowed. When the decision is undefined, system administrators should decide if the operation is allowed or denied by default.

The model supports operations for creating and deleting users, objects, subjects, attributes, policies and dependency relationships. While most of them only insert or remove one element from the corresponding sets, operations that delete resources also rely on composition dependency relationships. In addition to the specified resource and its attribute and policy assignments and dependency relationships, this operation also deletes all resources that are a part of that resource. $ResChildren_C$ function returns all resources that should be additionally deleted in one request.

## 4    Policy engine

In this section, we present the policy engine that supports RH-ABAC. It provides operations corresponding to model's administrative operations and authorization function. We developed it in order to show our model can be integrated with real-world systems. We optimized it for different use cases, so that a wider range of requirements is met. All optimizations can be turned on and off via configuration to achieve the desired behavior.

The policy engine is written in Go programming language. Clients can invoke any operation synchronously, using gRPC, or asynchronously, by sending requests to the NATS message broker. Asynchronous communication was enabled in order to decouple the engine from the rest of the system more easily. This property is especially important in microservice applications, which often use sagas [14] for transactive data updates in multiple services. To make data changes and response publishing atomic, we implemented the outbox pattern [15]. The message relay is implemented as a Go agent that periodically polls for outbox messages.

All data is currently stored in the Neo4j database, as it has a native graph data model facilitating the RDG traversal. Resources, attributes and policies are nodes in the graph, while dependency relationships and policy scopes are edges. There are two strategies for storing policy assignments. The first one persists only direct assignments. It creates edges from resources in the user scope to policies and from policies to resources in the object scope. For each authorization request, RDG must be traversed to find parents and applicable policies and policy priorities must be calculated.

The other strategy removes this overhead as it caches policies and their priorities. It does so by directly connecting policies with all users and objects for which that policy is applicable. The edge from user to policy stores the subject priority and the one from policy to object stores object priority. With this, authorization checks don't need to search potentially long paths to find parents or inherited policies. However, this introduces additional computation in administrative operations. The second strategy is recommended for systems with predominantly static resources and policy assignments, while systems with frequent updates may prefer the first strategy, as policy caching can degrade engine's performance.

When fast authorization is the first priority, the caching system can be configured. The engine uses multi-level cache where each engine instance has its own local cache and they all share a global Redis cache. The cache stores users, objects, their attributes and applicable policies for a particular operation. In the case of a cache hit, only the policy conditions have to be evaluated for the request. As the cache can retain stale data for some time, it should be used with caution.

## 5  Micro-cloud use case

Constellations (c12s)[1] is a proof-of-concept implementation of a model for dynamic formation of micro data centers, presented in [16, 17]. The model organizes nodes into disposable data centers in proximity to data sources, with the objective of offering edge computing as a service, like any other service in the cloud. To ensure resiliency and latency, the model consists of multiple layers: clusters, regions and topologies. Multiple nodes are organized into clusters. One or more clusters form a region. A topology comprises one or multiple regions.

The implementation supports the following operations:

- **Mutate** creates, updates and deletes topologies, regions and clusters;
- **List** displays the state of resources the user controls. They can request a high-level overview of all topologies for example, or a detailed view of clusters in a single region;
- **Query** shows free nodes that satisfy requirements set by the user;
- **Logs** returns user's activity history in the form of logs and traces;

To demonstrate feasibility of the RH-ABAC model, we extended c12s with access control support. When a new user registers, an organization in which he is granted all permissions is also created. Users can add other users to their organization. Every topology in the system belongs to one organization, created by some organization member. To facilitate policy administration, users can be assigned to groups.

Fig. 2 shows the transitive reduction of one possible RDG in the system. Users are nodes with the *u* prefix, while all other nodes are objects. Root resource is represented with the *root* node. Every other resource is dependent on it by composition relation. In this example, the system has one free node, *fnode:1*, and one organization, *org:o1*.

---

[1]  https://github.com/c12s

Users *u:u1* and *u:u2* are members of *org:o1*. User *u:u1* is a member of the group *g:g1* and user *u:u2* is a member of groups *g:g1* and *g:g2*. The organization owns one topology, *top:t1*, with two regions: *reg:r1* and *reg:r2*. Nodes *c:c1*, *c:c2*, *c:c3* and *c:c4* are clusters that form those regions and they consist of nodes *node:1, node:2, node:3* and *node:4*. Policy assignments are arrows whose start marks subject scopes and at their ends are object scopes. Policy assignments of policies *p1, p2* and *p3* and their properties are displayed in Table 2. Policy conditions are omitted since they function identically to the ones of other ABAC models.

The policy assignment of *p1* grants all users in the system the permission to query free nodes in order to form clusters, regions and topologies from them. This is because the subject scope is *root*, which is the parent of every user. Therefore, any user is in this subject scope. This is equivalent to other ABAC models, in which all policies are applicable to all subjects and objects.

The policy *p2* assignment allows all members of the *org:o1* to view information of every node in that organization. The mechanism of narrowing the subject and object scope, as is the case with this assignment, allows us to filter-out possibly many subjects and objects that are out of scope. For example, if we want to ask what users are granted this permission on *node:1*, we can instantly discard all users that are not members of the organization. This is of course true if there are no other policy assignments with the same permission and *node:1* in the object scope.

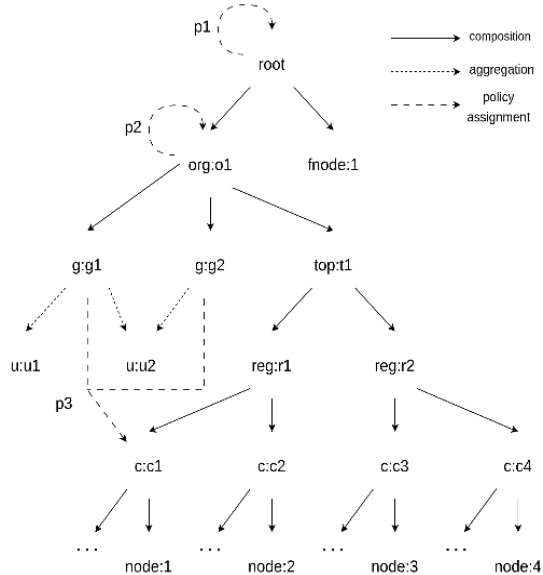

**Fig. 2.** Transitive reduction of the RDG for the micro-cloud use case

Authorization request for the subject of *u:u2*, object *node:1* and operation *node.get* evaluates two policies with conflicting decisions, *p2* and *p3*. While *p2* states that all members of the organization can view *node:1* information, policy *p3* denies the exe-cution of that operation for users that are members of both the group *g:g1* and *g:g2*. In this case, conflict resolution strategy must be enforced. Firstly, we must calculate the subject priority for both *p2* and *p3*, which are -2 and -1, respectively. Since *p3* has higher subject priority and a defined authorization decision, it will be kept, and *p2* won't. As there is only one policy left, we can skip other steps and return a response that operation execution is denied. However, if there were multiple policies left after the filtering, we would also have to calculate object priorities.

In this example, if a user issued a command to delete *top:t1* resource, apart from it, the regions *reg:r1* and *reg:r2* and all of their children resources would also be deleted. When a resource is being deleted, all of its attributed and dependency relationships should be removed, as well as policy assignments where that resource is a part of the subject or object scope, since they are not valid anymore. This cascading removal of resources requires only one operation for the entire hierarchy of resources, and it doesn't expect the client to know the structure of the hierarchy

**Table 2.** Policy assignment properties

| Policy | Operation | PermType | SubScope | ObjScope |
|---|---|---|---|---|
| *p1* | *freenode.list* | *allow* | *{root}* | *{root}* |
| *p2* | *node.get* | *allow* | *{org:o1}* | *{org:o1}* |
| *p3* | *node.get* | *deny* | *{g:g1, g:g2}* | *{c:c1}* |

## 6   Conclusion

In this paper, we demonstrated how user and object hierarchies lower the administrative overhead inherent to ABAC and proved hierarchical ABAC usability in distributed cloud environment. We achieved this by formally defining RH-ABAC model and implementing it in a policy engine. We outlined core RH-ABAC characteristics through a use case in micro clouds.

In order to have a thorough understanding of our model's strengths and weaknesses and explore its potential, we want to evaluate it on two levels. The first one involves comparing RH-ABAC expressive power and administration complexity to other hierarchical ABAC models, which is invaluable information when selecting an access control model. The second level is concerned with policy engine efficiency, as it is a major requirement for its integration with real-world systems. We want to quantify the trade-offs made by the two policy storing strategies, so that the most suitable one can be employed for a particular system.

## Acknowledgements

Funded by the European Union (TaRDIS, 101093006). Views and opinions expressed are however those of the author(s) only and do not necessarily reflect those of the European Union. Neither the European Union nor the granting authority can be held responsible for them.